# On the Ohmic-dominant heating mode of capacitively-coupled plasma inverted by boundary electron emission

Shu Zhang[1*], Guang-Yu Sun[2*], Jian Chen[3], Hao-Min Sun[4], An-Bang Sun[1†], Guan-Jun Zhang[1‡]

[1] *State Key Laboratory of Electrical Insulation and Power Equipment, Xi'an Jiaotong University, Xi'an, Shaanxi 710049, China*

[2] *Ecole Polytechnique Fédérale de Lausanne (EPFL), Swiss Plasma Center (SPC), CH-1015 Lausanne, Switzerland*

[3] *Sino-French Institute of Nuclear Engineering and Technology, Sun Yat-sen University, Zhuhai 519082. China*

[4] *Princeton Plasma Physics Laboratory, Princeton University, Princeton, New Jersey 08543, USA*

Electron emission from the boundary is ubiquitous in capacitively coupled plasma (CCP) and precipitates nonnegligible influences on the discharge properties. Here we present the PIC-MCC simulation of an Ohmic-dominant heating mode of capacitively coupled plasma where the stochastic heating vanishes and only Ohmic heating sustains the discharge, due to sheath inversion by boundary electron emission. The inverted CCP features negative sheath potential without Bohm presheath, hence excluding plasma heating due to sheath edge oscillation. The particle and energy transport of the proposed heating mode is analyzed. The influences of boundary electron emission flux, source voltage, and neutral pressure on the transition between classic and Ohmic-dominant CCP heating modes are shown with designated simulation scans. A modified inverse sheath-plasma coupling due to excessive ionization is discovered. In the end, key indicators of the proposed heating mode in plasma diagnostics are provided for future experimental verifications.

The energy balance of laboratory plasmas requires the choice of heating approaches to sustain the discharge,[1-3] among which the radio-frequency (RF) heating, either directly coupled to the plasma or through RF waves, is one of the most commonly employed heating mechanisms.[4, 5] The radio-frequency capacitively coupled plasma (CCP) discharge is established by applying RF current and voltage to an electrode immersed in the plasma, which plays a crucial role in plasma processing.[6-9] It is well known that the CCP features strong collisionless, stochastic heating at the sheath edge and Ohmic heating in bulk plasma.[10, 11] In this letter, we propose a special CCP heating mode without stochastic heating where Ohmic heating dominates in both sheath and plasma, achieved by strong boundary electron emission.

The influence of boundary electron emission on CCP discharge has been extensively studied in recent years, focusing on the secondary electron emission (SEE) induced by incident ions or electrons.[12-15] The SEE shifts the CCP from α-mode towards γ-mode as pressure increases, due to the higher ionization rate induced by the secondary electrons.[16] However, the

---

structure of Bohm presheath and capacitive sheath is retained. Since stochastic heating in CCP relies on the presheath density drop, it will not be affected by the boundary emission as long as Bohm criterion is valid. The presheath-sheath coupling can be modified when the boundary emission $\Gamma_{em}$ becomes sufficiently high such that it surpasses the plasma electron flux $\Gamma_{ep}$, i.e. the emission coefficient $\gamma = \frac{\Gamma_{em}}{\Gamma_{ep}} > 1$. Space-charge limited (SCL) sheath was supposed to be formed near floating or DC-biased boundary if $\gamma > \gamma_{crit} \approx 1$, with a local potential minimum called virtual cathode (VC) near the emissive surface.[17] Yet SCL sheath has been shown to be unstable if cold ions are born in VC due to charge exchange collision, and will be replaced by an inverse sheath featuring a wall potential well above the plasma.[18, 19] The inverse sheath has been proved to exist for floating, DC and RF plasma boundary in a variety of conditions, but it remains unclear how this special plasma-sheath coupling will affect the heating of RF-powered discharges, which will be the main purpose of the present work.[20-23]

A 1D3V PIC-MCC simulation code is employed to prove the existence of the Ohmic-dominant heating mode of CCP triggered by strong boundary electron emission. Employed code is updated from an existing CCP simulation program, which takes the velocity Verlet algorithm for particle integration, adaptive particle management as well as parallel computing using MPI into account. The code validity, convergence, and benchmark are presented in the author's previous works [16, 24, 25]. Firstly, the classic and Ohmic-dominant CCP heating modes are compared by showing distinct temporal and time-averaged profiles of sheath structure, plasma density, and power density distribution. The particle and power balance of the Ohmic-dominant heating mode free of stochastic heating is then analyzed. Subsequently, a series of parameter scans including applied voltage, boundary emission flux, and neutral pressure is performed to further reveal the approaches to achieve the mode transition. The effect of excessive ionization in high applied voltage and pressure range is discussed. Implications are given in the end to facilitate future experimental verification of the proposed heating mode.

The simulation studies a typical 1D3V parallel plate electrode structure, Fig. 1 (a). AC voltage source of 13.56 MHz is supplied on the power electrode, with a distance of 6.7 cm from the grounded electrode. Background neutral is 4 – 20 Pa Helium and the initial plasma density is $1.5 \times 10^{14}$ m$^{-3}$. Electrons are emitted from both electrode surfaces. The constant surface emission flux is adopted in the present simulation, corresponding to thermionic emission and photoemission of electrodes heated at constant temperature or exposed to radiation with constant intensity in practice. Secondary electron emission can also be adopted for highly emissive materials such as $SiO_2$ or Boron nitride.[26]

The discharge parameters of the classic (top row) and new (bottom row) CCP heating mode are shown in Fig. 1. In this article, the classic CCP is defined as the typical CCP with Debye sheath, and the new CCP is defined as the Ohmic-dominant heating CCP. RF voltage is 200 V for top row data without boundary emission and is 15 V for bottom row data. Electron flux of $5\times10^{18}$ $s^{-1}m^{-2}$ is emitted from both electrodes, corresponding to $\gamma$ of 0 ~ 5 at stable stage. The choice of different applied voltages will be discussed later on.

When strong boundary electron emission is supplied, the classic Debye sheath becomes inverted, with the mean electrode potential higher than the sheath edge potential, shown in Fig. 1 (b) (f). The formation of the inverse sheath with $\gamma > 1$ is consistent with existing simulation and experiments [18-21]. Note that the presheath is flat for an inverse sheath. The plasma densities are shown in Fig. 1 (c) (g). With strong emission, ion density decreases sharply in the inverse sheath as ions respond to time-averaged potential in the typical CCP frequency range. The high electron density in the inverse sheath is due to emitted electrons, most of which is reflected to the electrode before entering bulk plasma.

Electrons respond to real-time potential, and the temporal evolution of space potential is shown in Fig. 1 (d) (h). The oscillating sheath in classic CCP shields the external field and heats the electrons near the sheath edge just as balls collide on a moving solid wall. With strong emission, the external field is no longer shielded by plasma sheath and penetrates bulk plasma.[22]

The heating mechanism caused by sheath oscillation is referred to as stochastic heating, Fig. 1 (e), which requires the presheath potential (density) drop dictated by the Bohm criterion. The stochastic heating vanishes when the CCP is inverted since the presheath is flat and the Bohm criterion is invalid, Fig. 1 (i). In this case, Ohmic heating dominates the whole discharge region. The two spikes of power density near the edge are due to surface emitted electrons, which do not directly contribute to the power balance, to be shown later on.

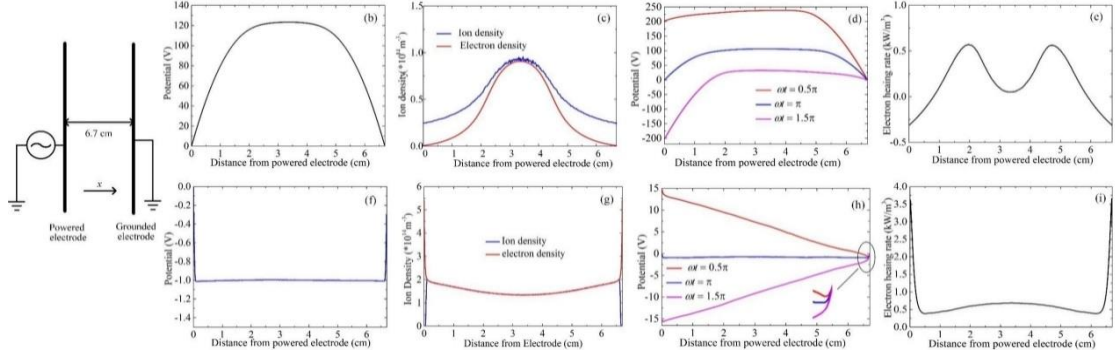

Fig. 1 (a) Schematic of the simulation model. (b)-(e): mean potential, mean plasma density, real-time potential, mean electron heating rate for CCP in classic heating mode. (f)-(i): same for CCP in Ohmic-dominant heating mode.

To explain the formation of the Ohmic-dominant heating mode, the particle and power balances are briefly shown below. The particle balance requires that the average net current to the electrode is zero and the ion flux is expressed by the ionization rate:

$$\Gamma_i = \langle \Gamma_{e,plasma} + \Gamma_{e,emitted} + \Gamma_{e,reflected} \rangle_T \qquad \text{Eq. 1}$$

$$2\Gamma_i = \int_0^x n_e n_n \langle \sigma_{ion} v_{e-n} \rangle \, dx \qquad \text{Eq. 2}$$

the four terms in Eq. 1 represent boundary fluxes of ion, plasma electron, emitted electron and reflected emitted electron, and the electron fluxes are averaged over a complete period. $\langle \sigma_{ion} v_{e-n} \rangle$ is the averaged rate coefficient and $n_e, n_n$ are electron and neutral densities.

In a CCP of classic heating mode, a considerable bulk ionization rate is required to compensate for the ion boundary loss, with ion flux dictated by the Bohm criterion. This is however no longer the case in the Ohmic-dominant heating mode since ion flux is mitigated by the mean inverse sheath, whose potential energy is comparable with emitted electron temperature, hence is well above the ion temperature.[22] Consequently, the particle balance of CCP in Ohmic-dominant heating mode is mainly controlled by the balance of three electron fluxes. The above analyses also imply that little ionization is needed to sustain the CCP discharge in Ohmic-dominant heating mode, whereas excessive ionization at high neutral pressures, as will be shown below, can modify the plasma-sheath coupling. This explains why high voltage is necessary for classic CCP simulation whereas CCP in Ohmic-dominant heating mode requires a much lower supply voltage.

The power balance of CCP in the Ohmic-dominant mode is significantly different from the classic heating mode, as shown in Fig.1. (e) (i). In classic heating mode, the sum of electron Ohmic heating power and stochastic heating power due to sheath oscillation is balanced by the energy losses at the boundary and e-n collision in plasma:

$$P_{edge,i} + P_{edge,ep} + P_{coll} = P_{ohmic} + P_{stoc} \qquad \text{Eq. 3}$$

here the boundary losses are contributed by both electron and ion flux ($P_{edge,ep}$ and $P_{edge,i}$). The e-n collision loss $P_{coll}$ mainly consists of elastic, inelastic collision, and ionization. $P_{stoc}$ mainly concentrates near the sheath edge and $P_{ohmic}$ is dominant in bulk plasma.

As mentioned above, stochastic heating does not exist in Ohmic-dominant heating mode CCP due to the absence of presheath density drop, leaving Ohmic heating as the only plasma electron heating source. The Ohmic heating in the Ohmic-dominant heating mode is more pronounced than the classic mode as it carries a larger field and current in the plasma. The Ohmic heating density is expressed as $S_e(x) = \langle J_{tot} * E \rangle_T$, with $J_{tot}$ the plasma current density and $E$ the electric field. In bulk plasma, the field is simply $V_{RF} cos(\omega t)/l$, leaving $S_{e,bulk} = \frac{1}{2}\sigma_{dc}\left(\frac{V_{RF}}{l}\right)^2$. Here the DC plasma conductivity is $\sigma_{dc} = \frac{ne^2}{mv_m}$ with $v_m$ the e-n collision rate. The electric field is much higher in the inverse sheath, leading to the two spikes of power density in Fig.1 (i).

Though the presence of surface-emitted electrons and reflected emitted electrons constitute the high electric field in the inverse sheath, they do not participate directly in the power balance. They are either reflected or lost from the other boundary, due to the long mean free path. The contribution of emitted electrons lies in the plasma electron acceleration. Combined above discussion, the power balance of CCP in Ohmic-dominant heating mode is shown below:

$$\frac{1}{2}\frac{ne^2}{mv_m}\left(\frac{V_{RF}}{l}\right)^2 l_{bulk} + P_{ohm,sheath} = 2\langle \Gamma_{ep}(\varepsilon_c + \varepsilon_{ep})\rangle_T \qquad \text{Eq. 4}$$

where the LHS represents the Ohmic heating in bulk plasma ($l_{bulk}$ the bulk plasma length excluding the sheath region) as well as plasma sheath ($P_{ohm, sheath}$). The RHS is the plasma electron energy loss at the boundary. $\Gamma_{ep}$ is the plasma electron loss flux at the boundary, $\varepsilon_c$ represents the energy loss due to electron-neutral collision (same as classic CCP) and $\varepsilon_{ep} = 2(T_{ep} + e\varphi_{inv})$ is mean electron energy at the boundary, $\varphi_{inv}$ is the real-time inverse sheath potential. In summary, the major difference between the two heating modes is that the Ohmic-dominant heating mode is only sustained by Ohmic heating and the plasma electron energy loss is significantly larger due to inverse sheath acceleration. Also, ion boundary loss is negligible when CCP is in Ohmic-dominant heating mode.

To further explore the transition between the two heating modes of CCP and understand how to sustain the Ohmic-dominant heating mode in practice, a series of parameter scans are performed below including the applied voltage, surface emission flux, and neutral pressure, shown in Fig. 2. The change of space potential distribution with increasing surface emission flux is shown in Fig. 2 (a) and (b). The absolute value of inverse sheath potential increases with emission flux which is consistent with existing inverse sheath theory.

When increasing applied voltage, the absolute value of the mean inverse sheath potential decreases while the peak potential of the plasma center gradually rises and surpasses the mean wall potential. It is generally assumed that the plasma potential should lie below the wall when an inverse sheath is formed in front of DC-biased or floating surfaces, due to strong boundary emission. This modified coupling between plasma and the inverse sheath is due to the high ionization rate at high applied voltage levels, where excessive ions are confined by the inverse sheath and accumulate in bulk plasma. This finally forms a potential drop from plasma to the boundary to accelerate ions so as to achieve particle balance dictated by Eq. 2. This phenomenon is not covered by the kinetic simulation of an inverted CCP where no realistic ionization is considered and the particle balance is artificially achieved by the source term.[22] The corresponding ionization rates and key potential indicators are shown in Fig. 2 (c) (d), supporting the above arguments. Note that the decreasing absolute value of mean inverse sheath potential with applied voltage which seemingly contradicts the previous theory[22] is due to changing plasma temperature, where all energy-related terms are normalized over plasma electron temperature in the inverse sheath theory. This further emphasizes the importance of power balance when analyzing the CCP modes.

Since the change of inverse sheath-plasma coupling is related to the high ionization rate, the similar effect also occurs when increasing neutral pressure, shown in Fig. 2 (e) (f). The ionization rate first increases and then decreases as the neutral pressure increases, while the potential drop in bulk plasma changes accordingly, consistent with the analyses above. The trend of ionization rate is due to lower electron temperature at high pressure values which dominates over increasing neutral density, as shown in Fig. 2 (g). Note that the inverse sheath structures and power density profiles with the potential hill in the bulk plasma due to excessive ionization are retained, indicating that CCP remains in Ohmic-dominant heating mode.

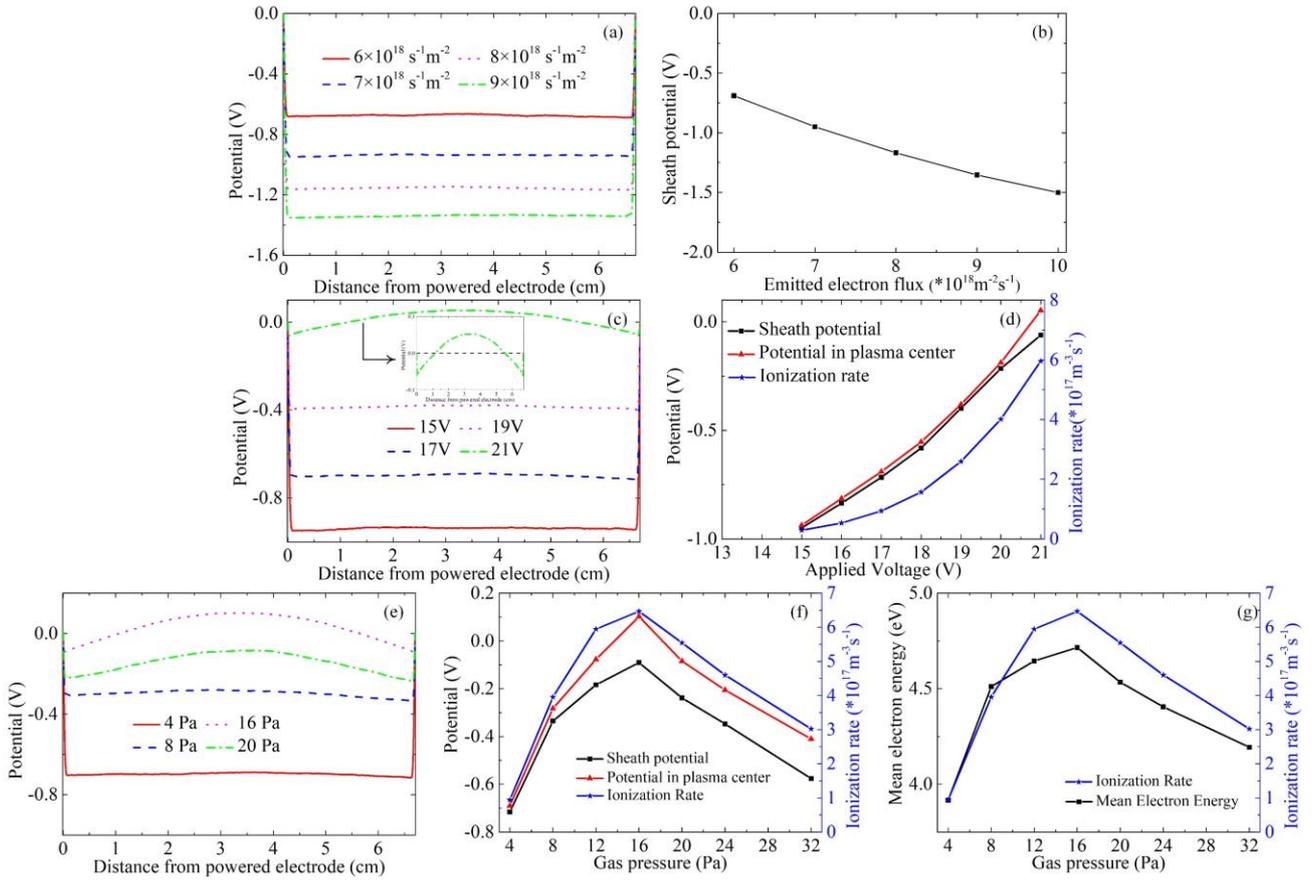

Fig. 2 (a) Sheath structure and (b) sheath potential with different emitted electron flux ($6\times10^{18} \sim 10^{19}$), where gas pressure (4 Pa) and applied RF voltage (15 V) are fixed; (c) sheath structure and (d) ionization rate under different applied voltages (13 ~ 21 V), with invariable emitted electron flux ($7\times10^{18}$ s$^{-1}$m$^{-2}$) and gas pressure (4 Pa); (e) sheath structure, (f) ionization rate and (g) mean electron energy under different gas pressure (4 ~ 20 Pa), with stabilizing the emitted electron flux ($7\times10^{18}$ s$^{-1}$m$^{-2}$) and applied voltage (17 V)

To verify the existence of the proposed heating mode in CCP, physical quantities such as potential, plasma density, or power density can be measured. The space potential profile near the inverse sheath is difficult to measure as the size of the sheath is comparable with the Debye length. The inverse sheath near the floating surface was demonstrated by Kraus and Raitses by comparing the potential difference between heated tungsten filament (emissive boundary) and the plasma.[21] The floating inverse sheath triggered by strong secondary electron emission was observed by Wang et al, by producing a wider sheath to facilitate the potential profile measurement.[27] Indirect interpretation of floating inverse sheath using I-V traces of the emissive probe was adopted by Chi-Shung et al.[28] The measurement of potential distribution in RF condition is supposed to be similar. In addition, boundary electron density higher than bulk density can also justify the proposed plasma heating mode. The power density requires the current and electric field measurement and high resolution near the inverse sheath could be a concern.

In conclusion, the Ohmic-dominant CCP heating mode is proposed and compared with the classic CCP heating mode with theory and simulation. The Ohmic-dominant heating mode contains no stochastic heating as the Bohm criterion is not valid for plasma coupled with the inverse sheath. The power density distribution for CCP in Ohmic-dominant heating mode is approximately flat in the bulk plasma due to Ohmic heating and features two spikes near the boundary, due to emitted electrons. Contrary to the classic CCP where an intense bulk ionization rate is needed to compensate for the ion boundary loss, little ionization is needed when CCP is in Ohmic-dominant heating mode, exhibiting unique particle and power balance. Scan of applied voltage and neutral pressure suggests that excessive ionization leads to a potential drop from the bulk plasma to the inverse sheath edge, which accelerates ions to overcome the inverse sheath barrier and achieve particle balance. Key indicators of the proposed heating mode are provided for experimental verifications.

## ACKNOWLEDGMENTS

The authors would like to thank Dr. Han Jia in Swiss Plasma Center for the fruitful discussions. This research was conducted under the auspices of the National Natural Science Foundation of China (NSFC, Grants No. 51827809, No. U1766218 and No. 52077169) and the National Key R&D Program of China (Grant No. 2020YFC2201100).

## DATA AVAILABILITY STATEMENTS

The data that support the findings of this study are available from the corresponding author upon reasonable request.